\documentclass[prc,aps,amsfonts,amsmath,superscriptaddress,floatfix,twocolumn]{revtex4}
\usepackage{graphicx,float} 
\usepackage{epsfig} 
\usepackage{rotating}
\usepackage{mwe}    
\usepackage{subfig}

\begin{document}

\title{Microscopic description of $\alpha$, $2\alpha$, and cluster decays of $^{216-220}$Rn and $^{220-224}$Ra}
\author{J. Zhao}
\affiliation{Center for Circuits and Systems, Peng Cheng Laboratory, Shenzhen 518055, China}
\author{J.-P. Ebran}
\affiliation{Universit\'e Paris-Saclay, CEA, Laboratoire Mati\`ere en Conditions Extr\^emes, 91680, Bruy\`eres-le-Ch\^atel, France}
\author{L. Heitz}
\affiliation{IJCLab, Universit\'e Paris-Saclay, CNRS/IN2P3, 91405 Orsay Cedex, France}
\author{E. Khan}
\affiliation{IJCLab, Universit\'e Paris-Saclay, CNRS/IN2P3, 91405 Orsay Cedex, France}
\author{F. Mercier}
\affiliation{IJCLab, Universit\'e Paris-Saclay, CNRS/IN2P3, 91405 Orsay Cedex, France}
\author{T. Nik\v si\' c}
\affiliation{Physics Department, Faculty of Science, University of Zagreb, 10000 Zagreb, Croatia}
\author{D. Vretenar}
\affiliation{Physics Department, Faculty of Science, University of Zagreb, 10000 Zagreb, Croatia}

\begin{abstract} 
$\alpha$ and cluster decays are analyzed for heavy nuclei located above $^{208}$Pb on the chart of nuclides:  $^{216-220}$Rn and $^{220-224}$Ra, that are also candidates for observing the $2 \alpha$ decay mode.
A microscopic theoretical approach based on relativistic Energy Density Functionals (EDF), is 
used to compute axially-symmetric deformation energy surfaces as functions of quadrupole, octupole and hexadecupole collective coordinates. Dynamical least-action paths for specific decay modes are calculated on the corresponding potential energy surfaces. The effective collective inertia is determined using the perturbative cranking  approximation, and zero-point and rotational energy corrections are included in the model. 
The predicted half-lives for $\alpha$-decay are within one order of magnitude of the experimental values. 
In the case of single $\alpha$ emission, the nuclei considered in the present study exhibit least-action paths that differ significantly up to  the scission point. The differences in $\alpha$-decay lifetimes are not only driven by Q values, but also by variances of the least-action paths prior to scission.  In contrast, the $2 \alpha$ decay mode presents very similar paths from equilibrium to scission, and the differences in lifetimes are mainly driven by the corresponding Q values. The predicted  $^{14}$C cluster decay half-lives are within three orders of magnitudes of the empirical values, and point to a much more complex pattern compared to the $\alpha$-decay mode. 
\end{abstract}
 
\date{\today}

\maketitle

\section{\label{intro} Introduction}

In a series of recent studies, a microscopic approach based on relativistic Energy Density Functionals (EDF) has successfully been applied to the phenomenon of $\alpha$ radioactivity. In this framework, self-consistent calculations of potential energy surfaces (PES), as functions of various deformation degrees of freedom, are used to predict the least-action dynamical path leading to a particular decay. Using the Wentzel-Kramers-Brillouin (WKB) approximation \cite{del10}, $\alpha$-decay half-lives have been analyzed in $^{104}$Te, $^{108}$Xn, $^{212}$Po and $^{224}$Ra nuclei \cite{mer20,mer21}. The relativistic EDF-based microscopic framework used to model $\alpha$ decay, has also been applied to the description of spontaneous and induced fission \cite{zha16,tao17,zha19,zha19a,zha20,zha21,zha22,ren22,ren22a}. 

This approach has recently been used to predict a new mode of decay, where two $\alpha$-particles are simultaneously emitted back-to-back \cite{mer21}. The new $2 \alpha$ decay mode should be verified experimentally, which raises the question of the optimal candidate nucleus that must have: i) the shortest $2 \alpha$ decay lifetime, ii) the largest branching ratio, and iii) realistic production rates at present-day exotic beam facilities. The most obvious candidate is $^{216}$Rn, which corresponds to 2 $\alpha$ particles added to the doubly magic $^{208}$Pb. Neighboring nuclei, such as $^{218}$Rn, and their $\alpha$-decay `parent nuclei', for instance $^{220,222}$Ra, could also be considered.

From the theoretical point of view, it is also relevant to investigate whether the present approach is able to describe cluster decay, which is known to occur in this area of the nuclear chart. Since the relativistic EDF-based method can be used to compute both fission and $\alpha$-decay observables, the intermediate case, cluster emission, should be investigated as well. This will also provide a link with previous calculations of cluster decay based on the Gogny EDF, performed in a similar framework \cite{war11}. It should be noted, however, that the relativistic EDF-based model appears to be the only fully microscopic approach able to describe $\alpha$-decay in a quantitative agreement with available data. This feature could be related to the ability of relativistic EDFs to describe the formation of $\alpha$ cluster states in nuclei  \cite{ebr12,ebr13,ebr14,ebr14a,zha15a,zho16,mar18,mar19}. Especially, the relation between the formation of $\alpha$ clusters states and $\alpha$ decay has been explored in Refs. \cite{ebr18,ebr21}.

The aim of the present work is threefold: i) to assess the quality of the microscopic relativistic EDF-based calculations of $\alpha$ emission lifetimes, by comparing with empirical values for the nuclei of interest, namely $^{216-220}$Rn and $^{220-224}$Ra, as well as $^{212}$Po; ii) to explore whether the relativistic EDF approach could also predict cluster decay, and the level of agreement with data, as well as possible open questions raised by a universal microscopic description of fission, cluster, $2 \alpha$ and $\alpha$ decays; and iii) to analyze possible $2 \alpha$ decays of nuclei in the region ${\rm ^{208}Pb}+2 \alpha = {\rm ^{216}Rn}$, which should be favored as a 2 $\alpha$ emitter candidate. On this last point, $^{216}$Rn, but also its $\alpha$-parent nucleus, as well as slightly more neutron-rich nuclei, will be analyzed. The aim is to identify optimal nuclei that also have large production rates, usually from a primary reaction on a Uranium target.

The present study is based on calculations performed in the Relativistic Hartree-Bogoliubov (RHB) framework. For a detailed review, we refer the reader to Refs.~\cite{vre05,men16,ebr19}. Self-consistent calculations of deformation energy surfaces are performed using the DD-PC1~\cite{nik08} relativistic functional and a separable finite-range  pairing interaction. The potential energy surfaces (PES) are calculated using quadrupole, octupole and hexadecupole deformations as collective degrees of freedom. Section \ref{sec2} describes the formalism and the details of the calculation. Predictions of $\alpha$-decay half-lives for $^{216-220}$Rn and $^{220-224}$Ra are discussed in section \ref{sec3}. Section \ref{sec4} explores the symmetric $2 \alpha$ decay mode, and in section \ref{sec5} we analyze $^{14}$C cluster emission from $^{222}$Ra and $^{224}$Ra. A brief summary of the principal results is included in section \ref{sec6}.

\section{\label{sec2} Relativistic Hartree-Bogoliubov calculations of decay rates}

The process of decay by an $\alpha$-particle or cluster emission is modelled along a dynamical path $L$, determined by minimizing the action integral~\cite{bra72,led73}:

\begin{equation}
\label{eq:act_integration}
S(L) = \int_{s_{\rm in}}^{s_{\rm out}} {1\over\hbar} 
  \sqrt{ 2\mathcal{M}_{\rm eff}(s) \left[ V_{\rm eff}(s)-E_0 \right] } ds ,
\end{equation}
where $\mathcal{M}_{\rm eff}(s)$ and $V_{\rm eff}(s)$ are the effective 
collective inertia and potential, respectively.
$E_0$ is the ground-state energy of the nucleus, and the integration
limits correspond to the classical inner ($s_{\rm in}$) and outer turning points ($s_{\rm out})$, defined by $V_{\rm eff}(s) = E_0$.

The effective inertia is computed from the multidimensional inertia tensor $\mathcal{M}$ \cite{bra72,bar78,bar81,sad13,sad14}
\begin{equation}
\mathcal{M}_{\rm eff}(s) = \sum_{ij} \mathcal{M}_{ij} {dq_i \over ds} {dq_j \over ds}\;,
\end{equation}
where $q_i(s)$ denotes the collective coordinate as a function of the path's length. The collective inertia tensor is calculated from the self-consistent and deformation-constrained RHB solutions for the quasiparticle wave functions and energies, using the adiabatic time-dependent Hartree-Fock-Bogoliubov (ATDHFB)  method \cite{bar11}. In the perturbative cranking approximation, the collective inertia reads \cite{zha15}

 \begin{equation}
\label{eq:pmass}
\mathcal{M} = \hbar^2 {\it M}_{(1)}^{-1} {\it M}_{(3)} {\it M}_{(1)}^{-1}, 
\end{equation}
 
 where 
\begin{equation}
\label{eq:mmatrix}
\left[ {\it M}_{(k)} \right]_{ij} = \sum_{\mu\nu} 
    {\left\langle 0 \left| \hat{Q}_i \right| \mu\nu \right\rangle
     \left\langle \mu\nu \left| \hat{Q}_j \right| 0 \right\rangle
     \over (E_\mu + E_\nu)^k}\; . 
\end{equation}
$|\mu\nu\rangle$ are two-quasiparticle 
states, and $E_\mu$ and $E_\nu$ the corresponding quasiparticle energies. $\hat{Q}_i$
denotes the multipole operators that represent collective degrees of freedom.
The effective collective potential $V_{\rm eff}$ is obtained by subtracting the 
vibrational zero-point energy (ZPE) from the total RHB 
deformation energy. Following the prescription of 
Refs.~\cite{sta13,bar07,sad13,sad14}, 
the ZPE is computed using the Gaussian overlap approximation, 
\begin{equation}
\label{eq:zpe}
E_{\rm ZPE} = {1\over4} {\rm Tr} \left[ {\it M}_{(2)}^{-1} {\it M}_{(1)} \right].
\end{equation}
In addition, the rotational energy correction (REC) is also subtracted
\begin{equation} 
\label{eq:rec}
E_{\rm REC} = \frac{<J^2>}{2\mathcal{I}_Y}\;,
\end{equation}
where the expectation values of the angular momentum operator are evaluated with respect to the self-consistent and deformation-constrained RHB states, and $\mathcal{I}_Y$ is the Yoccoz moment of inertia \cite{eg04} for a given multipole operator $\hat{Q}_i$:
\begin{equation}
\label{eq:yoc}
\mathcal{I}_Y (\hat{Q}_i) = \sum_{\mu\nu} 
    { 2\left(\left|\left\langle 0 \left| \hat{Q}_i \right| \mu\nu \right\rangle\right|^2\right)^2
       \over \left| \left\langle 0 \left|\hat{Q}_i \right| \mu\nu \right\rangle\right|^2 (E_\mu + E_\nu)}\; . 
\end{equation}
The REC approximation of Eq.~(\ref{eq:rec}) is known to be valid for large deformations \cite{war11}, and replaces full angular momentum projection which, in the case of heavy nuclei, is generally computationally prohibitive.

It should be noted that the REC was not taken into account in our previous calculations of $\alpha$ decay based on  relativistic EDFs \cite{mer20,mer21}. Therefore, the present results should be, in principle at least, more accurate. A typical effect of this correction is to decrease the lifetimes of $\alpha$-decay by about one order of magnitude. For instance, in the case of $^{224}$Ra, the inclusion of the REC leads to a decrease of the $\alpha$ emission lifetime from 9.5 d to 5.7 d. We have verified that the rotational energy correction globally improves the results in comparison to data. Therefore, in the present work, all calculations are performed by including the REC. 

The microscopic self-consistent solutions of the deformation-constrained RHB equations, that is, the 
single-quasiparticle energies and wave functions on the entire energy surface as functions of the 
quadrupole, octupole, and/or hexadecapole deformations, provide the microscopic input for computing the 
collective inertia, zero-point energy and rotational energy corrections.

In practice, the least-action path is built from the inner turning point to the scission point, whose position is determined by 
monitoring the integrated density distribution, that is, the density distribution of the emerging fragment.  
As scission point, we select the point with emerging fragment
mass equal to the mass of the $\alpha$ particle, $2 \alpha$ particles, or $^{14}$C cluster. For the single $\alpha$ or cluster emission, beyond scission the configuration with two well separated fragments becomes the lowest energy solution, and the energy curve up to the outer turning point $s_{\rm out}$ can be approximated by the classical expression for two uniformly charged spheres: 
\begin{equation}
\label{eq:coulomb}
V_{\rm eff}(\beta_{3}) = e^2 {Z_{1}Z_{2} \over R} - Q,
\end{equation}
where $R$ represents the distance between the centers of mass of the fragments, and 
the second term is the experimental $Q$ value.
We use Eqs. (9) and (10) of Ref. \cite{war11} to approximate the relation between $R$ and the octuple moment $Q_{30}$,
\begin{equation}
\label{eq:R-Q}
Q_{30} = f_3 R^3,
\end{equation}
with
\begin{equation}
\label{eq:f3}
f_3 = {A_1 A_2 \over A} {(A_1 - A_2) \over A},
\end{equation} 
and $\beta_{30}=4\pi Q_{30} / 3AR^3$.
The corresponding effective collective mass reads
\begin{equation}
\label{eq:cmass}
\mathcal{M}_{\rm eff} = {\mu \over 9Q^{4/3}_{30} f^{2/3}_3}, 
\end{equation}
where $\mu=m_{n} A_1A_2/(A_1+A_2)$ is the reduced mass of the two fragments, and 
$m_{n}$ denotes the nucleon mass \cite{war11}. 

In the case of emission of two $\alpha$ particles, to calculate the contribution of the action from the scission point to the outer turning point $s_{\rm out}$, we consider the superposition of each $\alpha$-plus-nucleus Coulomb interaction, namely:
\begin{equation}
\label{eq:coulomb2}
V_{\rm eff}(\beta_{2}) = 2e^2 {Z_{1}Z_{2} \over R} - Q_{2\alpha},
\end{equation}
where $R$ represents the distance between the centers of mass of the fragments (the index 1 refers to the heavy fragment, and 2 to the $\alpha$-particle). The approximate relation between R and the quadrupole moment is 
\begin{equation}
\label{eq:R-Q2}
Q_{20} = 2A_2 R^{2}\;.
\end{equation}
The corresponding effective collective mass reads
\begin{equation}
\label{eq:cmass2}
\mathcal{M}_{\rm eff} = {\mu \over 8A_2 Q_{20}}, 
\end{equation}
where $\mu=m_{n} A_2/2$ is the reduced mass of the (2$\alpha$+ heavy) fragments.

Therefore, the path involved in the action integral of Eq. (\ref{eq:act_integration}) consists of the least-action path 
from $s_{\rm in}$ to scission, and the energy is approximated by the Coulomb potential from scission to $s_{\rm out}$ \cite{war11}. 
The decay half-life is calculated as $T_{1/2}=\ln2/(nP)$, where $n$ is the
number of assaults on the potential barrier per unit 
time~\cite{bar78,bar81,sad13,sad14}, 
and $P$ is the barrier penetration probability in the WKB approximation
\begin{equation}
P = {1 \over 1+\exp[2S(L)]}. 
\label{prob}
\end{equation}
We choose $E_{0} = 1$ MeV in Eq. (\ref{eq:act_integration}) for the value of the collective ground state energy.
For the vibrational frequency $\hbar\omega=1$ MeV, the corresponding value of $n$ is $10^{20.38}$ s$^{-1}$.

In the present study, for the single $\alpha$ emission the collective space is 3-dimensional $(\beta_{20},\beta_{30},\beta_{40})$, while for the symmetric 2$\alpha$ emission, because of reflection symmetry, the collective space can be built from the coordinates $\beta_{20}$ and $\beta_{40}$. In the case of $^{14}$C emission, two models will be analyzed:
i) a 2D $(\beta_{20},\beta_{30})$ calculation, where the scission point is determined by a discontinuity in $\beta_{40}$, to be compared with previous Gogny EDF-based results \cite{war11} and, in the case of $^{222}$Ra ii) a full 3D  $(\beta_{20},\beta_{30},\beta_{40})$ calculation, in which scission is determined as a point where the integrated density of the emitted fragment corresponds to 14 nucleons.  

We note that the present deformation-constrained calculations are performed using a method with linear constraints that has successfully been applied to fission (see Ref.~\cite{zha15} for details). The Dirac-Hartree-Bogoliubov equations are solved by expanding the nucleon spinors in the basis of a 3D harmonic oscillator, with 20 oscillators shells. In addition to the particle-hole channel determined by the choice of the EDF, a separable pairing interaction of finite-range \cite{dug04,tia09} is used that reproduces the pairing gap in nuclear matter as calculated with the D1S parametrization of the Gogny force~\cite{ber91,tia09}. As in Ref. \cite{mer21}, we have fine-tuned the neutron and proton pairing strengths to reproduce the
empirical pairing gaps of the isotope $^{224}$Ra. Compared to the original values, this corresponds to an increase of the neutron and proton pairing strengths by 9\% and 12\%, respectively. This modification is consistent with the conclusions of the recent global study of the separable pairing interaction, when used with relativistic energy density functionals \cite{wa13,af13}. 
We have used the modified pairing strength parameters for all the Ra and Rn nuclei considered in the present work.

\section{\label{sec3} $\alpha$-decay}

\begin{figure}[tbh!]
\scalebox{0.206}{\includegraphics{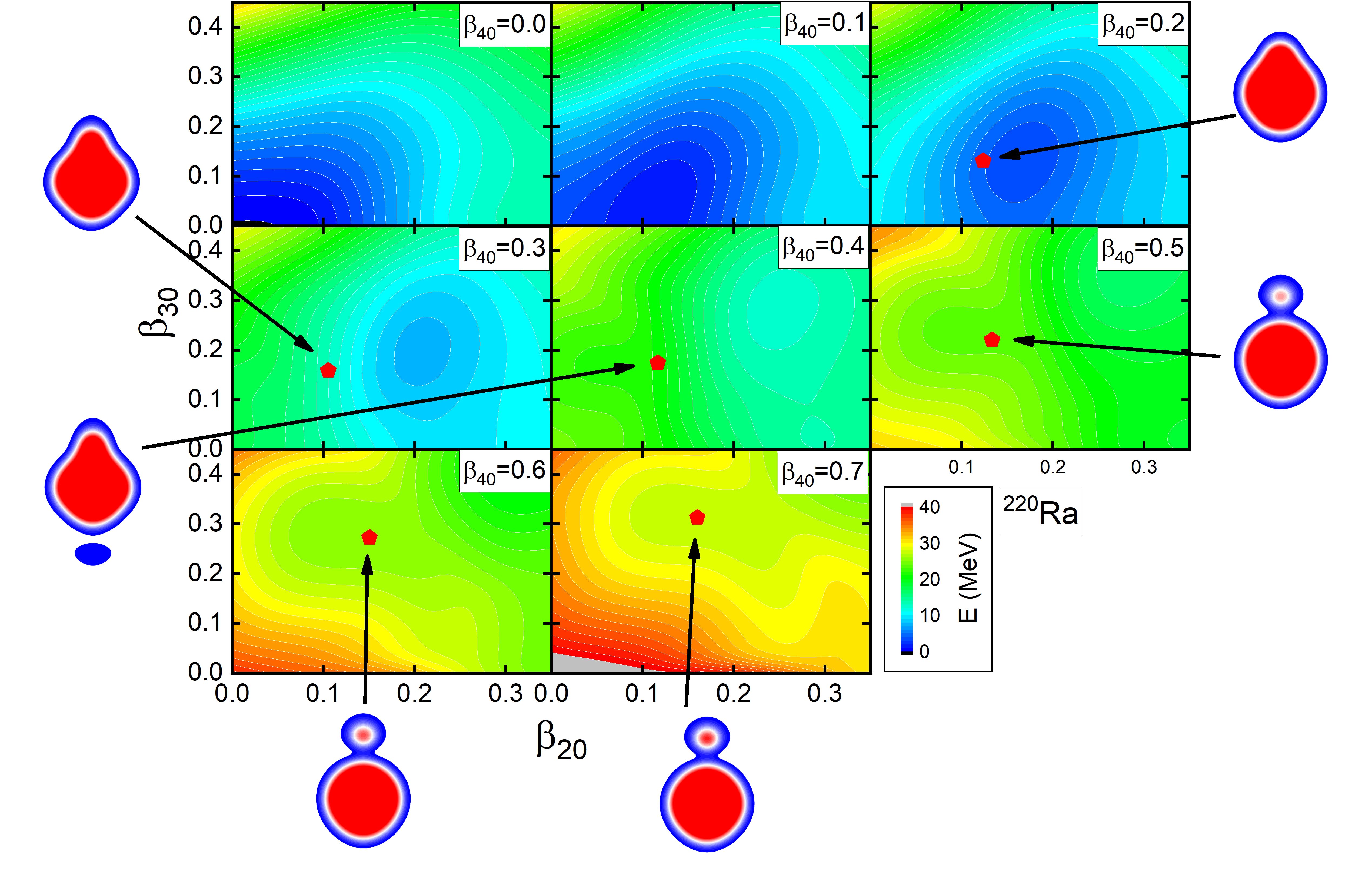}}
 \caption{Deformation energy surface of $^{220}$Ra in the
axially-symmetric quadrupole-octupole plane, for selected
values of the hexadecapole deformation $\beta_{40}$. Calculations
are performed using the RHB model based on the DD-PC1
functional and a separable pairing interaction. Contours join
points on the surface with the same energy, and red circles
indicate the points on the dynamical (least-action) path for $\alpha$ emission.
The insets display the intrinsic nucleon densities at selected points on the dynamical path.}    
 \label{fig:220Raalpha}
\end{figure}

In the case of $\alpha$ decay, the importance of including at least three deformation degrees of freedom has been shown in our previous calculations \cite{mer20,mer21}. The quadrupole degree of freedom corresponds to the elongation of the nucleus, while octupole deformations are essential to model the formation of an $\alpha$ particle on the surface of a nucleus. The formation of a neck between the $\alpha$ particle and the daughter nucleus before emission, gives rise to a more pronounced hexadecapole deformation. For the $\alpha$ decay of $^{220}$Ra, figure \ref{fig:220Raalpha} displays several projections of the PES in the axially symmetric quadrupole-octupole plane, for different values of the hexadecapole deformation parameter $\beta_{40}$. The least-action path remains close to the energy minimum of the deformation surfaces, with some deviations  due to the effect of the collective inertia. Along the path from the inner ($s_{\rm in}$)  turning point to the scission point, the value of $\beta_{20}$ increases slightly from 0.12 to 0.16. The increase of $\beta_{30}$ is more pronounced, from 0.12 to 0.32, and even more so for $\beta_{40}$ from 0.2 to 0.7. Hence, in the case of $\alpha$ decay from $^{220}$Ra, the major deformation degrees of freedom at work are the octupole and hexadecapole ones. The scission point on the ($\beta_{20}$, $\beta_{30}$, $\beta_{40}$) surface is found at ($0.16, 0.31, 0.68$), and the least-action integral up to this point has a value of 8.10, whereas the one from scission to the outer turning point ($s_{\rm out})$ is 14.18. The predicted half-life is 60 ms, to be compared with the experimental value of 18 ms.
 
It is interesting to compare the predictions for the half-lives of $\alpha$-decay of $^{220}$Ra and $^{212}$Po. In the latter case, the scission point is very close to the one of $^{220}$Ra: (0.16,0.33,0.70). The least-action integral calculated form the inner turning point to scission is 6.58, and 9.46 from scission to the outer turning point. We note that the action up to the scission point is smaller than in the case of $^{220}$Ra. This is due to the almost spherical shape of the equilibrium minimum of $^{212}$Po: the least-action path starts from a small $\beta_{20}$ value, close to zero, because of the vicinity of the doubly magic nucleus $^{208}$Pb. The difference between the calculated least-action integrals is then magnified by the well-known Q-value effect after the scission point, making the total action much smaller in the case of $^{212}$Po. 
About 25\% of the difference in the total action between $^{220}$Ra and  $^{212}$Po is due to nuclear dynamics before the scission point, rather than to the Q-value effect. The correct description of the dynamical path before the scission point is crucial, as demonstrated by the predicted half-life: 0.2 $\mu$s for $^{212}$Po, to be compared to the experimental value of 0.3 $\mu$s. 

\begin{figure}[tbh!]
\scalebox{0.25}{\includegraphics{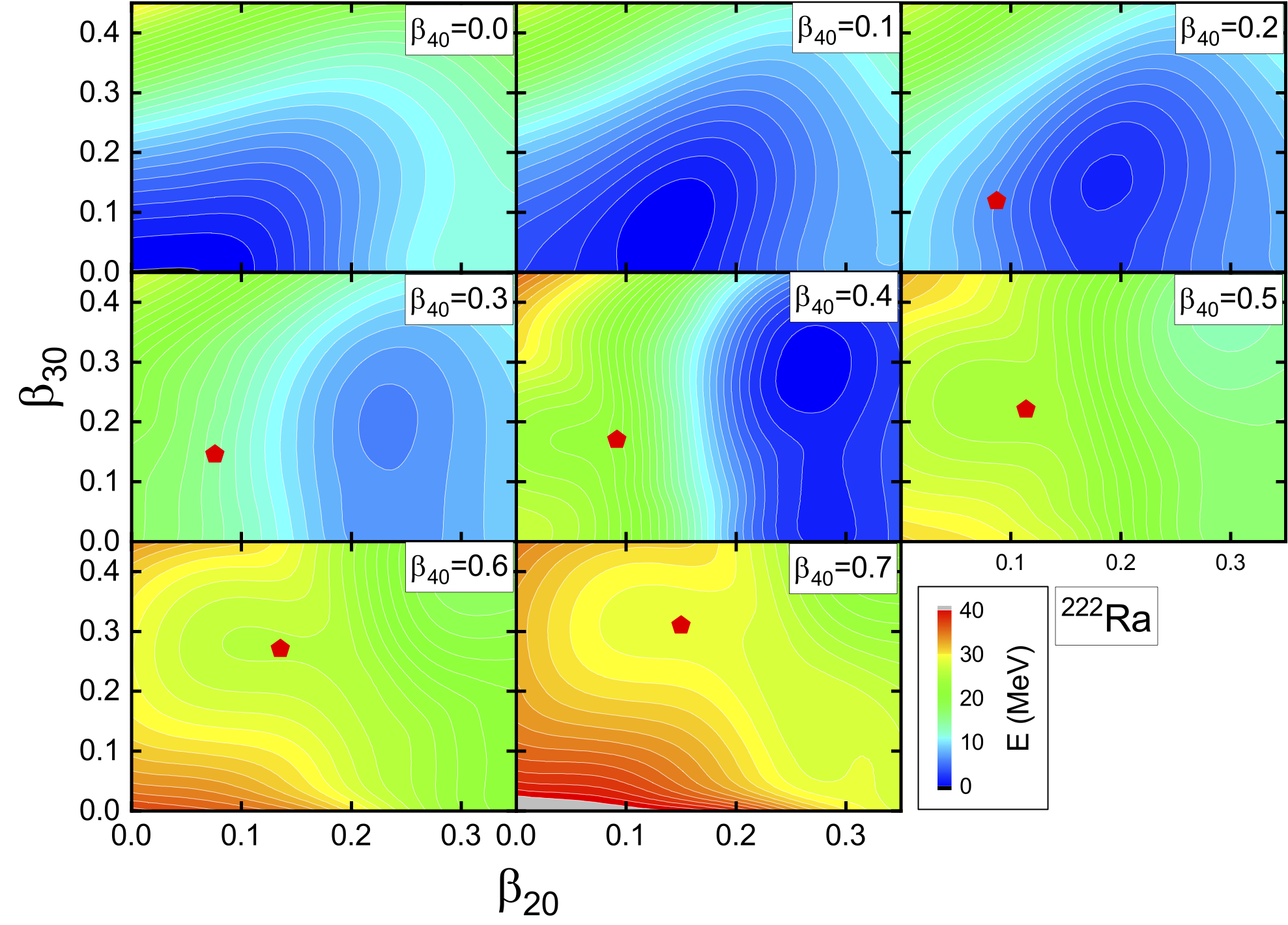}}
 \caption{Same as in the caption to Fig. \ref{fig:220Raalpha} but for $^{222}$Ra}
 \label{fig:222Raalpha}
\end{figure}

Figure \ref{fig:222Raalpha} illustrates the calculation of the $\alpha$-decay of $^{222}$Ra. The evolution of the path is similar to the one of $^{220}$Ra. The scission point is obtained at ($0.15, 0.31, 0.68$) on the ($\beta_{20}$, $\beta_{30}$, $\beta_{40}$) surface, which is almost identical to the case of $^{220}$Ra. The least-action integral up to the scission point is 9.03, and the one in the outer region starting from this point is 17.02. The corresponding half-life is 122 s, in comparison to the experimental value 38 s. Again, a correct description of the least-action path prior to the scission point is crucial to predict a realistic half-life within one order of magnitude of the empirical value. 

\begin{figure}[tbh!]
\scalebox{0.25}{\includegraphics{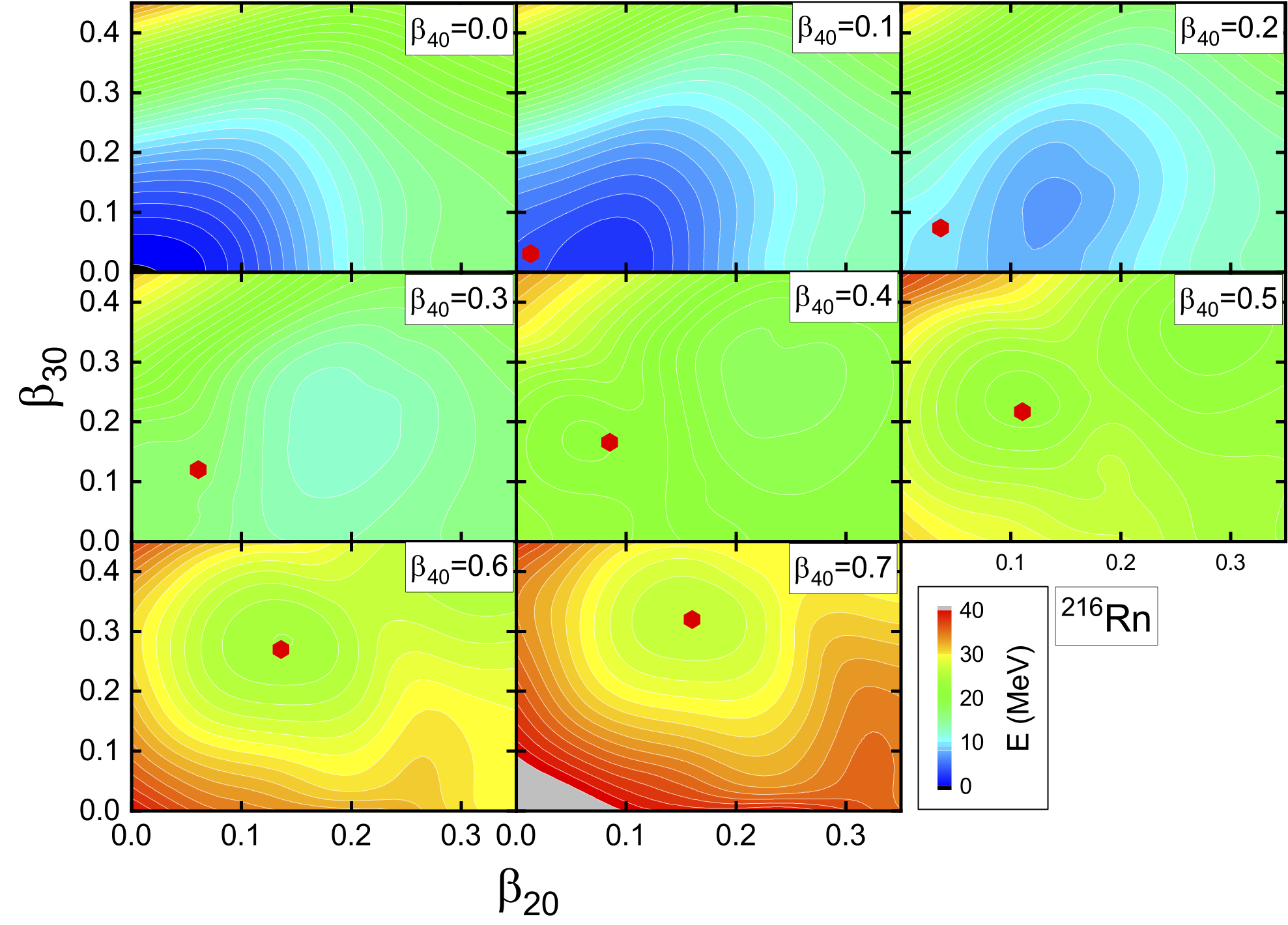}}
 \caption{Same as in the caption to Fig. \ref{fig:220Raalpha} but for $^{216}$Rn}
 \label{fig:216Rnalpha}
\end{figure}

\begin{figure}[tbh!]
\scalebox{0.25}{\includegraphics{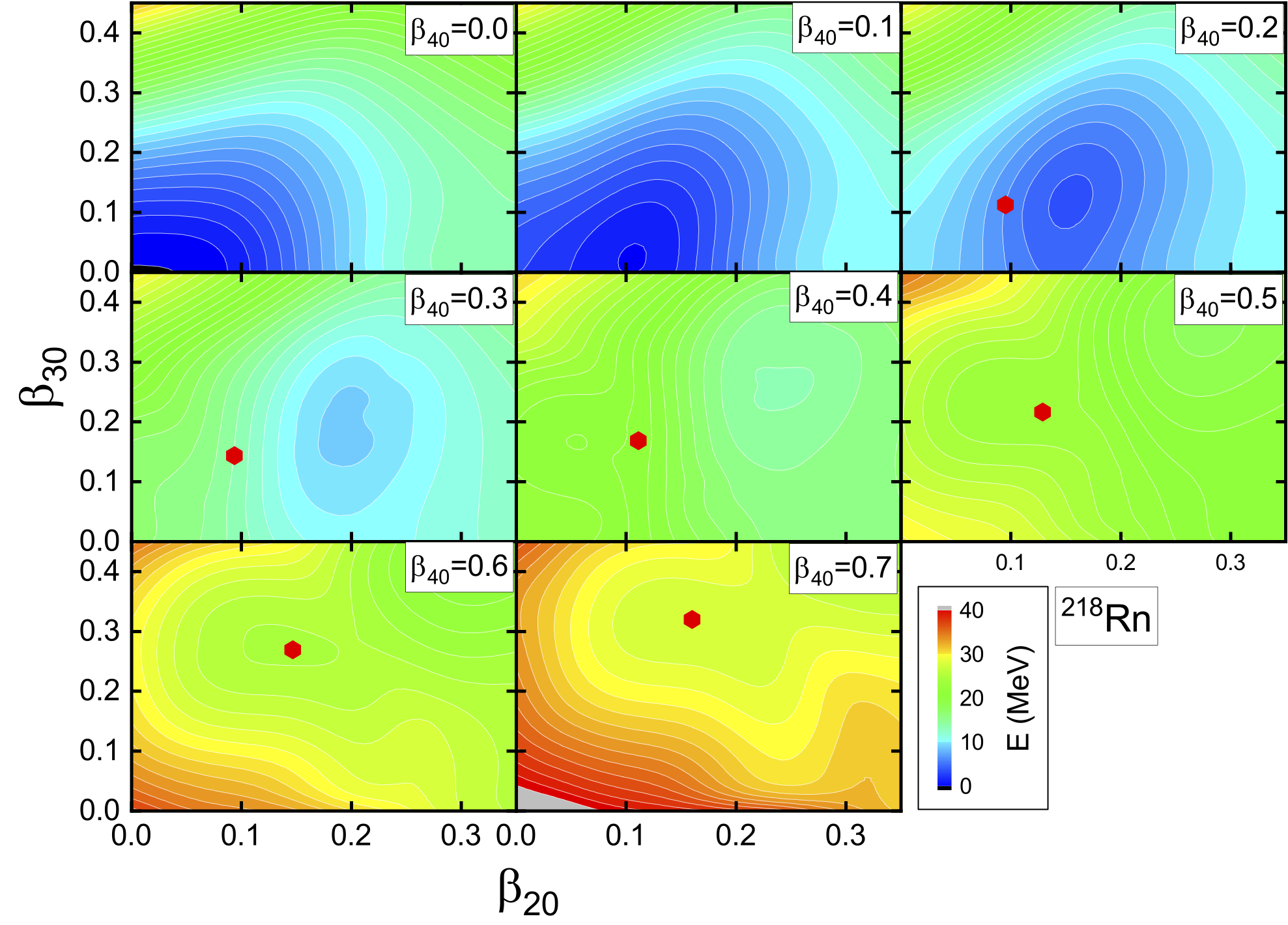}}
 \caption{Same as in the caption to Fig. \ref{fig:220Raalpha} but for $^{218}$Rn} 
 \label{fig:218Rnalpha}
\end{figure}

For the $\alpha$-decay of $^{216}$Rn and $^{218}$Rn, in 
figures  \ref{fig:216Rnalpha} and \ref{fig:218Rnalpha}, respectively, we display the corresponding projections of the 3D deformation energy surfaces. The case of $^{216}$Rn
is similar to the one of $^{212}$Po, with a starting point of the least-action path close to sphericity. The scission point  is found at ($0.16, 0.32, 0.70$) on the ($\beta_{20}$, $\beta_{30}$, $\beta_{40}$) surface, the action integral up to the scission point is 8.12, and the one in the outer region 11.83. The predicted half-life is 185 $\mu$s, and the experimental one is 45 $\mu$s.
For $^{218}$Rn, the PES is comparable to those of the Ra nuclei discussed above. The scission point is at ($0.16, 0.32, 0.70$), the action up to the scission point is 8.58, and the one from this point is 14.32. Here, despite the difference in the dynamical paths before the scission point (cf. Figs. \ref{fig:216Rnalpha} and \ref{fig:218Rnalpha}), the calculated actions before the scission point are close, though a bit larger in the case of $^{218}$Rn. After this point, because of the smaller Q-value, the action is also larger in the case of $^{218}$Rn. The calculated half-life 35ms is in agreement with the experimental value.

The case of $^{220}$Rn is close to the $^{218}$Rn one, with similar PES and small variations on the dynamical path. The scission point is at ($0.12, 0.32, 0.70$), the action up to the scission point is 8.03, and the one from this point is 17.48. This larger value, compared to the previous  $^{216,218}$Rn cases, is due to a smaller Q-value. The predicted half-life is 40.8 s, and the experimental one is 55.6 s.

Table \ref{tab:hl} summarizes the $\alpha$ decay half-lives calculated using the present method. All the theoretical values are of the same order of magnitude as the experimental ones. This result demonstrates the validity of the microscopic approach, which explicitly takes into account various important effects in a consistent framework: pairing, collective inertia, energy corrections, etc. It should also be noted that there are no free parameters specifically adjusted to reproduce the empirical half-lives. The only input is a universal energy density functional, and the pairing interaction. 

\setlength{\tabcolsep}{10pt}
\begin{center}
\begin{table}[h]
\caption{\label{tab:hl}}
Experimental and theoretical values for the $\alpha$-emission half-lives, predicted 2$\alpha$ symmetric emission half-lives, and the corresponding branching ratios. BR denotes the log$_{10}$ of the branching ratio.
\begin{tabular}{cccccc}
  & T$_{exp}$ & T$_{\alpha}$  &log$_{10}$ (T$_{2\alpha}$[s]) & BR
 \\ \hline
 $^{212}$Po & 0.3 $\mu$s    &  0.2  $\mu$s   &  17.23 & -23.8
\\ 
 $^{216}$Rn & 45 $\mu$s&   185 $\mu$s   &  2.67 &  -6.4
\\
 $^{218}$Rn & 35 ms & 35 ms  & 5.06 &  -6.5
 \\ 
 $^{220}$Rn & 55.6 s & 40.8 s  & 8.04 &  -6.4
 \\ 
 $^{220}$Ra & 18 ms &   60 ms   &  6.1 &  -7.3
\\ 
 $^{222}$Ra & 38 s &  122  s &  9.23 &  -7.1
\\ 
 $^{224}$Ra & 3.6 d &  5.7 d &  13.03 &  -7.3
 \\
\end{tabular}
\end{table}
\end{center}
\setlength{\tabcolsep}{6pt}

\section{\label{sec4} Symmetric two-$\alpha$ decay}

\begin{figure}[tbh!]
\scalebox{0.32}{\includegraphics{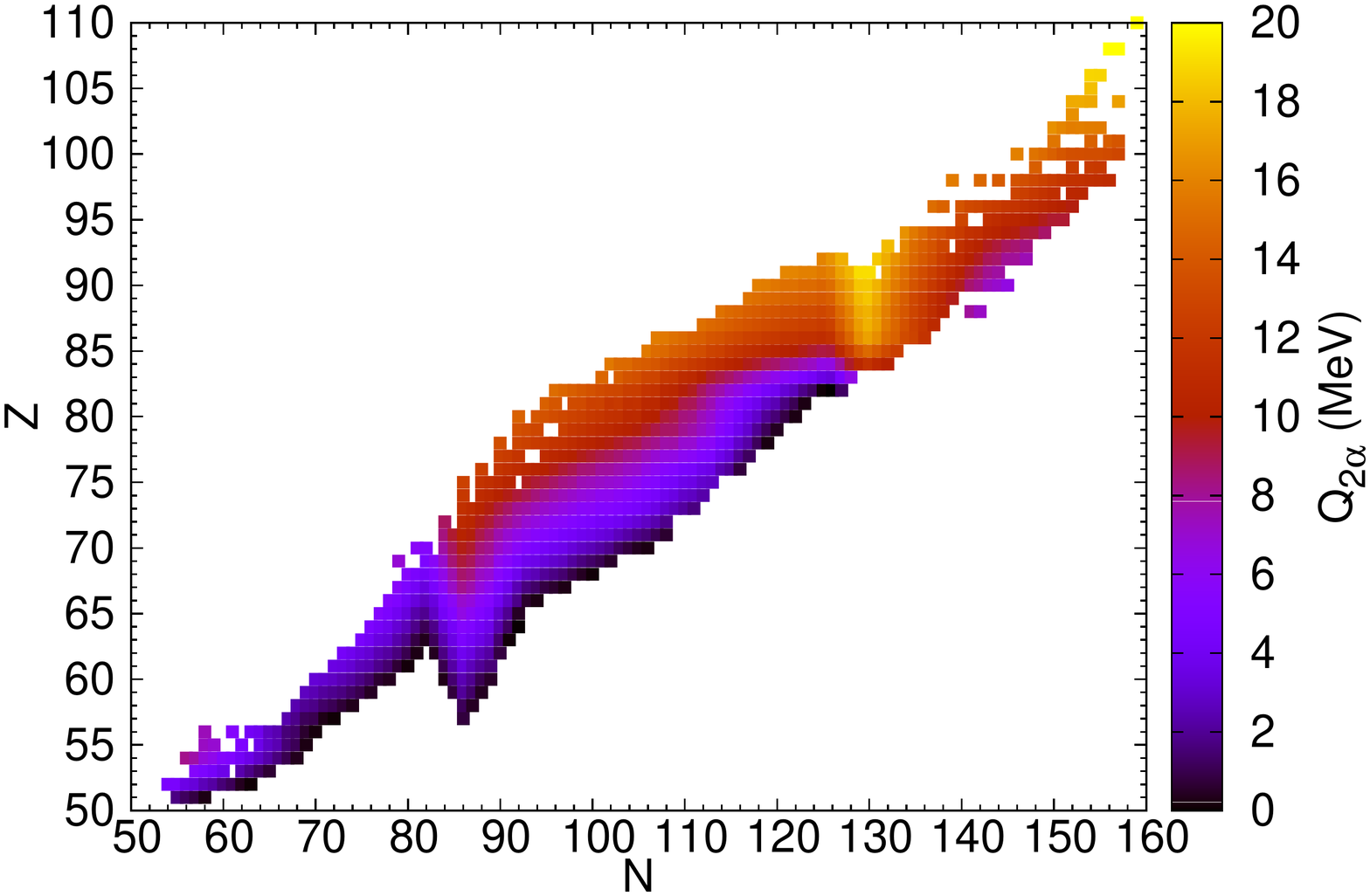}}
 \caption{$N>50$ and $Z>50$ nuclides with positive Q$_{2\alpha}$ values.}
 \label{fig:chartQ2a}
\end{figure}

The simultaneous symmetric emission of two $\alpha$ particles is a new mode of decay which has been predicted recently \cite{mer21}. In this process, the two $\alpha$ particles are emitted back to back, and the least-action path is calculated on the surface of quadrupole and hexadecapole deformations. The octupole degree of freedom plays no role here. Considering the condition of positive 
Q$_{2\alpha}$ value, almost all of the one-$\alpha$ emitter nuclei are predicted to be also 2-$\alpha$ emitters, with a few exceptions \cite{mer21}. However, the branching ratio for this mode is generally very small, about 10$^{-8}$ in $^{224}$Ra. This is nevertheless of the same order of magnitude as the branching ratios for 
cluster radioactivity, which has already been detected. Therefore, it is necessary to focus on the best 2 $\alpha$-decay candidates for a possible measurement. 
Figure \ref{fig:chartQ2a} displays the Q$_{2\alpha}$ values on the nuclear chart, for nuclei exhibiting positive Q$_{2\alpha}$ values, where Q$_{2\alpha}$ is defined by:
\begin{equation}
Q_{2\alpha}=Q_{\alpha1}+Q_{\alpha2}+\Delta E\;.
\end{equation}
Q$_{\alpha1(2)}$ are the Q-values of the daughter and grand-daughter nuclei for single $\alpha$ decay, determined from experimental masses \cite{aud03}. $\Delta$E is the difference between the sum of the excitation energies of the daughter and grand-daughter nuclei in the sequence of single $\alpha$ decays, and the excitation energy of the grand-daughter nucleus in the $2 \alpha$ decay. We consider, for simplicity, $\alpha$ or $2 \alpha$ transitions involving only the ground states of the daughter or grand-daughter nuclei (that is, $\Delta E$=0). Considering the empirical Q$_{2\alpha}$ values in Fig. \ref{fig:chartQ2a}, it appears that the most favored region for $2 \alpha$ decay is located in a narrow strip north-west of $^{208}$Pb. The optimal experimental candidate should be as close as possible to this region, with an intensity of production large enough to detect a few possible $2 \alpha$ decays. Unstable nuclei around this region are best produced
from a Uranium primary beam. Taking into account these constraints, some of the best candidates could be $^{216}$Rn, which corresponds to $^{208}$Pb+2$\alpha$, followed by $^{218}$Rn, $^{220}$Ra, $^{222}$Ra, and $^{224}$Ra. We note that nuclei around $^{218}$Th or $^{222}$U are even more favorable when considering 
Q$_{2\alpha}$ values. However, their current production rate is too low to perform dedicated experiments on the $2 \alpha$ decay mode.

\begin{figure}[tbh!]
\scalebox{0.30}{\includegraphics{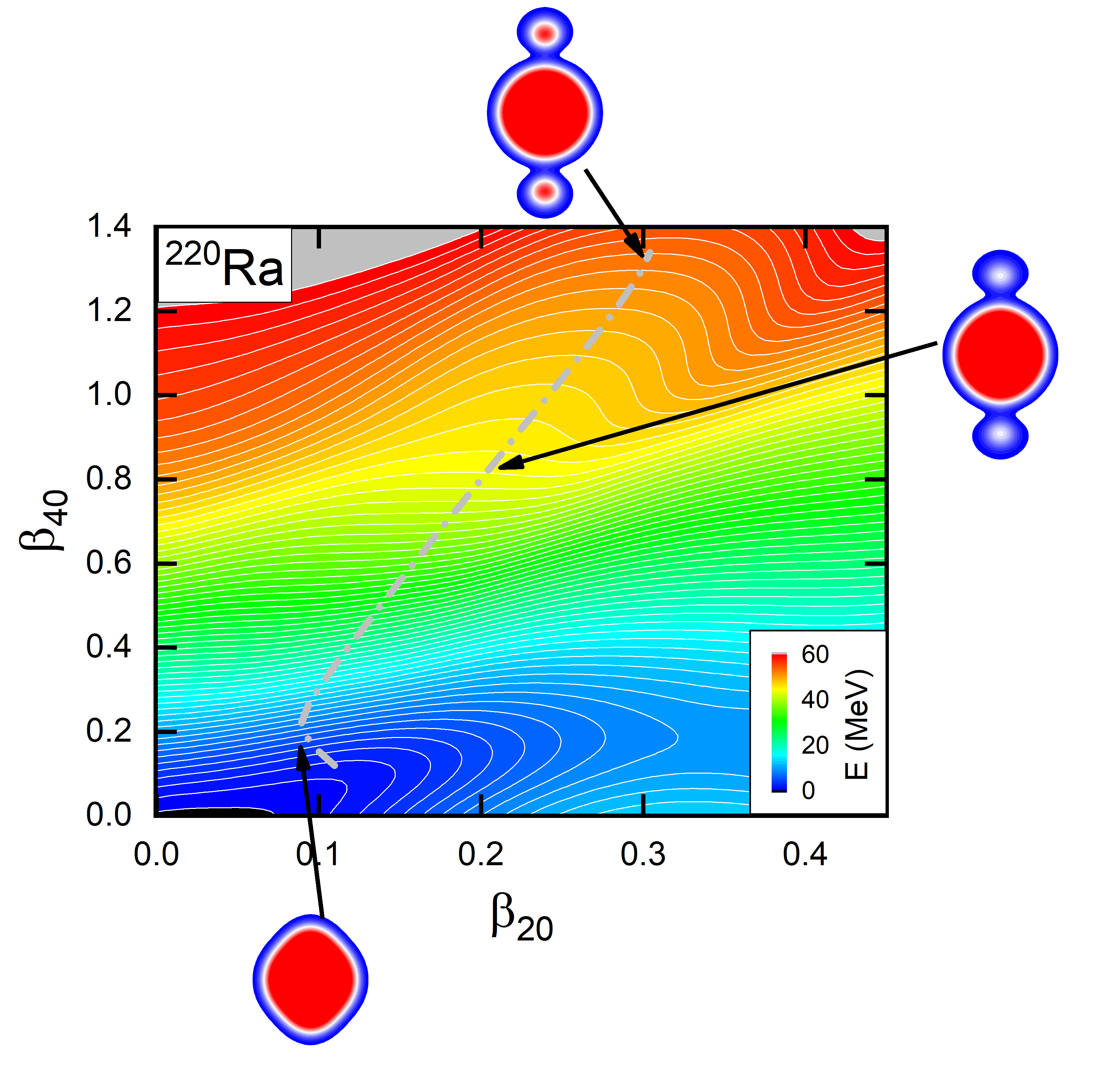}}
 \caption{Reflection symmetric deformation energy surface
of $^{220}$Ra  in the quadrupole-hexadecapole axially-symmetric
plane. Calculations have been performed using the RHB model
based on the DD-PC1 functional, and a separable pairing interaction.
The grey curve denotes the dynamical
(least-action) path for 2$\alpha$ emission from the equilibrium
deformation to scission.
The insets display the intrinsic nucleon densities at selected points on the dynamical path.}
 \label{fig:220Ra2alpha}
\end{figure}

\begin{figure}[tbh!]
\scalebox{0.30}{\includegraphics{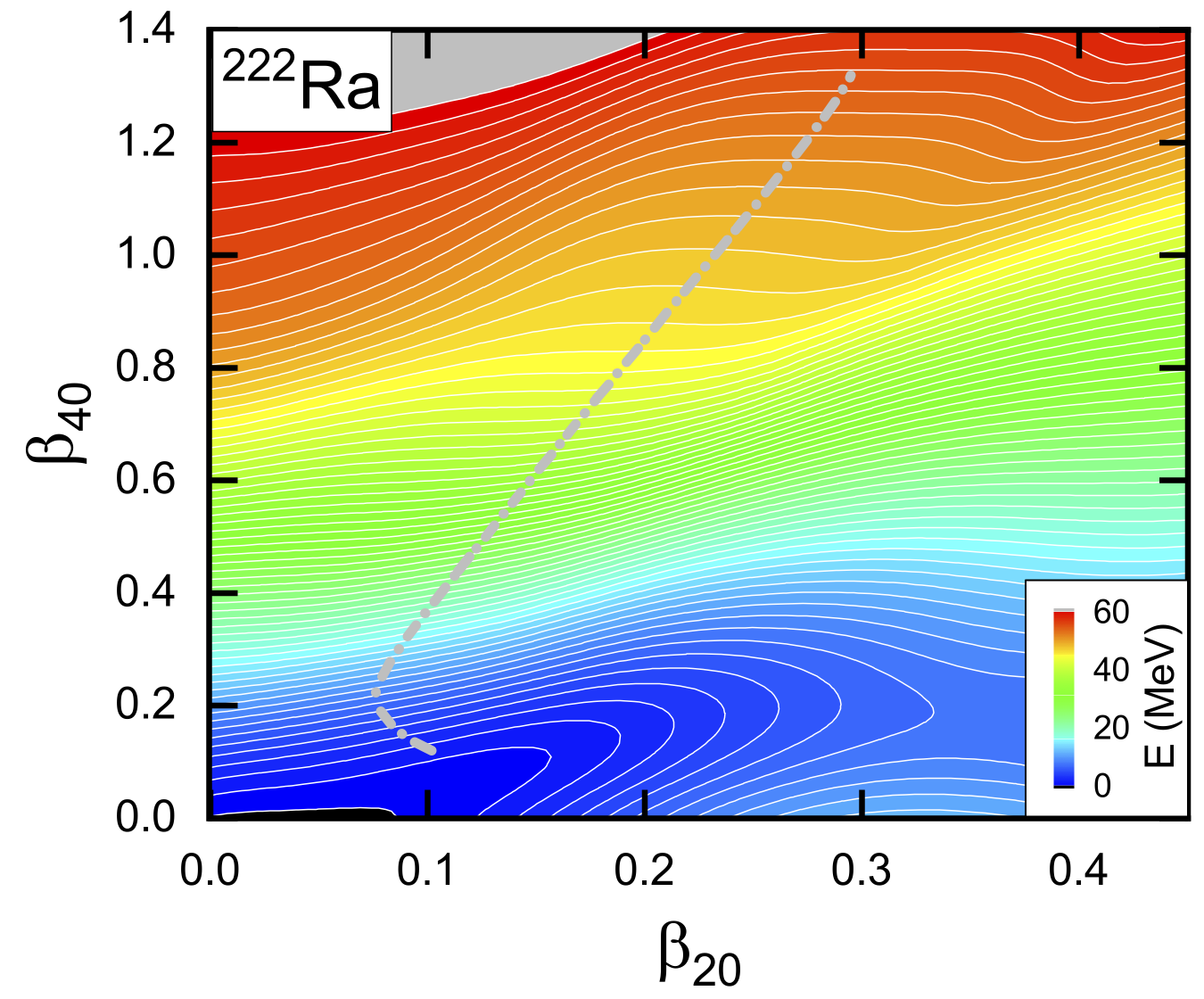}}
 \caption{Same as in the caption to Fig. \ref{fig:220Ra2alpha} but for $^{222}$Ra.}
 \label{fig:222Ra2alpha}
\end{figure}

\begin{figure}[tbh!]
\scalebox{0.30}{\includegraphics{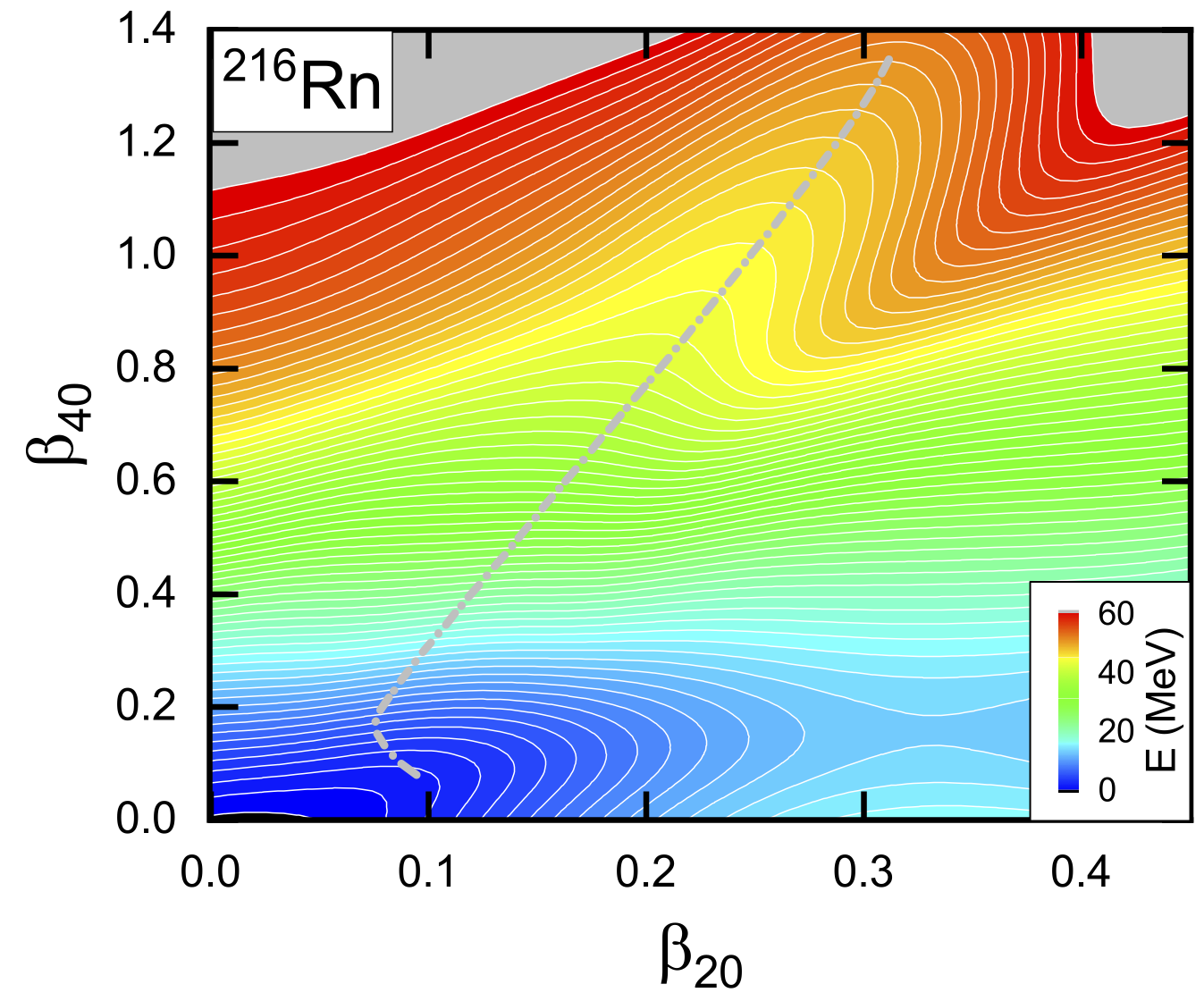}}
 \caption{Same as in the caption to Fig. \ref{fig:220Ra2alpha} but for $^{216}$Rn.}
 \label{fig:216Rn2alpha}
\end{figure}

\begin{figure}[tbh!]
\scalebox{0.30}{\includegraphics{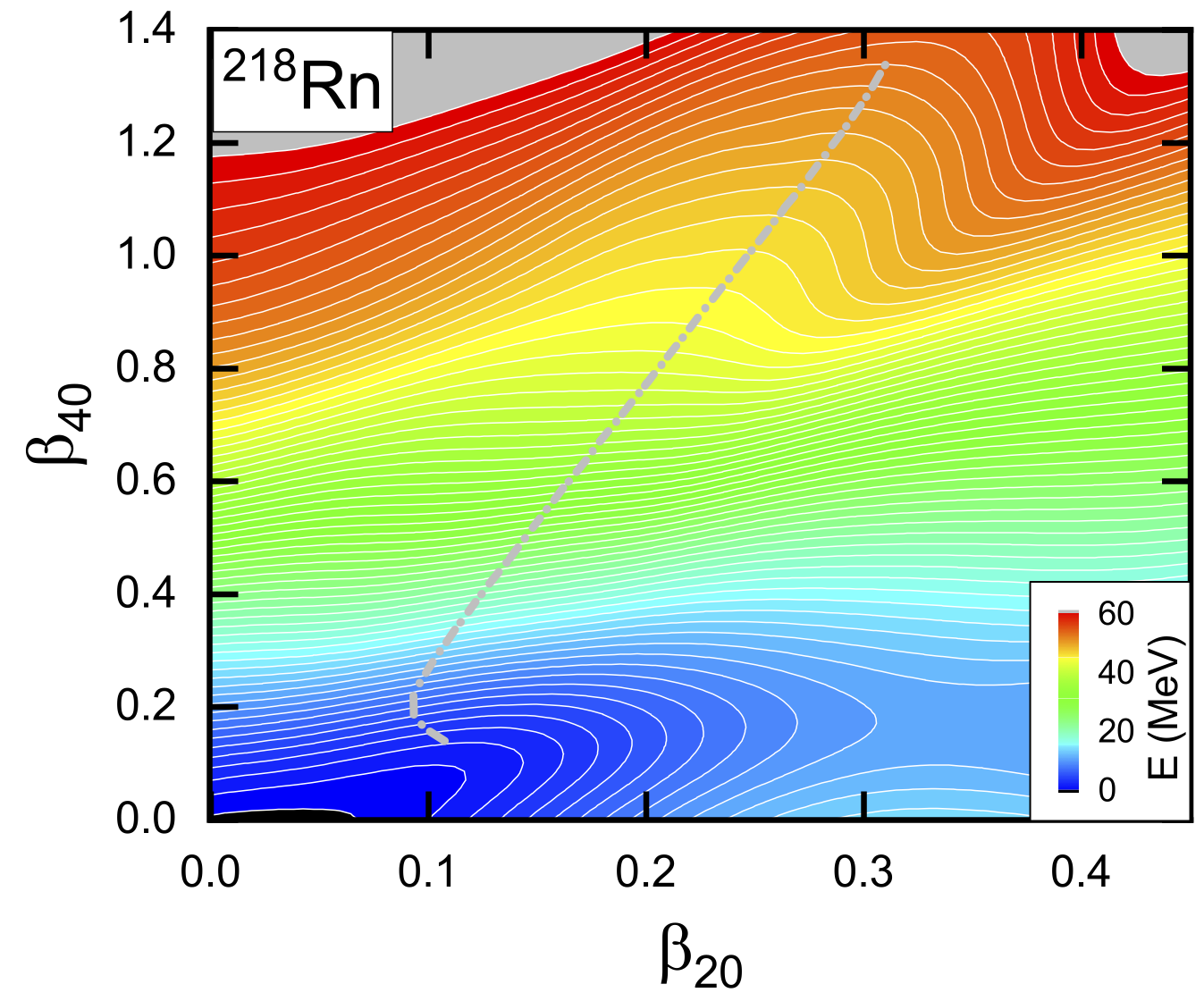}}
 \caption{Same as in the caption to Fig. \ref{fig:220Ra2alpha} but for $^{218}$Rn.}
 \label{fig:218Rn2alpha}
\end{figure}

In Figs. \ref{fig:220Ra2alpha}-\ref{fig:218Rn2alpha} we show the RHB deformation energy surfaces in the quadrupole-hexadecapole plane for the nuclei $^{220,222}$Ra and $^{216,218}$Rn. In these
four cases, the least-action paths for the symmetric $2 \alpha$ emission are very similar: a slight initial decrease of 
$\beta_{20}$, followed by a linear increase of both $\beta_{20}$ and $\beta_{40}$, leading to very large values of the hexadecapole deformation parameter at scission. The scission point is located
at $\beta_{20} = 0.30$ and $\beta_{40} = 1.34$ in the case of  $^{220}$Ra, and almost the same for the other three nuclei (less than 2\% change in the deformation parameters). Correspondingly, 
the least-action integrals up to the scission point are also rather similar: 13.93, 14.36, 12.70, and 12.61 for $^{220,222}$Ra and $^{216,218}$Rn, respectively. 
However, it is the Q$_{2\alpha}$ values that play the main role in differentiating the $2 \alpha$ decay rates in these nuclei. The least-action integrals from the scission point to the outer turning points are: 16.76, 19.92, 14.02, and 16.87 for $^{220,222}$Ra and $^{216,218}$Rn, respectively. The corresponding half-lives, listed in Table \ref{tab:hl}, can directly be correlated to these values of the action integral: the shortest half-life is calculated for $^{216}$Rn, whereas the longest one is for $^{222}$Ra. The above conclusions also apply to the case of $^{220}$Rn with an action before the scission point of 13.01, the one after the scission point being 19.90.

The predicted $2 \alpha$ decay branching ratios (BR) listed in Table \ref{tab:hl}, show that the most unfavorable case for this mode appears to be $^{212}$Po. This could be understood because of the $^{208}$Pb+1$\alpha$ configuration of this nucleus. In contrast, $^{216}$Rn, which corresponds to $^{208}$Pb+2$\alpha$, is predicted to be the optimal case for simultaneous symmetric $2 \alpha$ decay.
However, neighboring nuclei such as $^{218,220}$Rn or $^{220-224}$Ra also exhibit BRs that do not differ by more than an order of magnitude. This is due to the simultaneous increase of the action integrals in both the $\alpha$ and $2 \alpha$ channels, when compared to $^{216}$Rn. In summary, $^{216-220}$Rn and $^{220-224}$Ra should be considered as candidates for an experimental search of the $2 \alpha$ decay mode.

It would also be interesting to consider additional geometric degrees of freedom, such as triaxiality, but this is beyond the scope of this work, and hardly feasible from a computational point of view. However, the inclusion of additional degrees of freedom would generally decrease the action calculated along the decay path and, hence, the predicted half-lives for the two-$\alpha$ decay. The half-lives calculated in the present study should be considered as upper limits for the two-$\alpha$ decay mode.  

Three-body effects are also relevant for two-particle, e.g., two-proton decay. However, it is difficult to draw a direct analogy between two-proton and two-$\alpha$ decays, because preformation of $\alpha$ clusters has a large impact on the decay probability which, of course, is not the case for two-proton decay. In the present calculation, all possible interactions are considered before the scission point, because the least action path is calculated starting from nucleonic degree of freedom, displaying the emergence of the two $\alpha$ along this path. After the scission point, each $\alpha$-nucleus interaction is taken into account, while the interaction between the two $\alpha$ is neglected. This corresponds to a simplified case of a full three-body treatment. Translated in the description of two-proton decay, this would rather correspond to the so-called direct emission model \cite{pfu12}, which yields a correct order of magnitude for the decay half-lives \cite{pfu13}, although only the full three-body treatment provides the correct momentum density distribution of the emitted protons. 

\section{\label{sec5} Cluster decay}

\begin{figure}[tbh!]
\scalebox{0.30}{\includegraphics{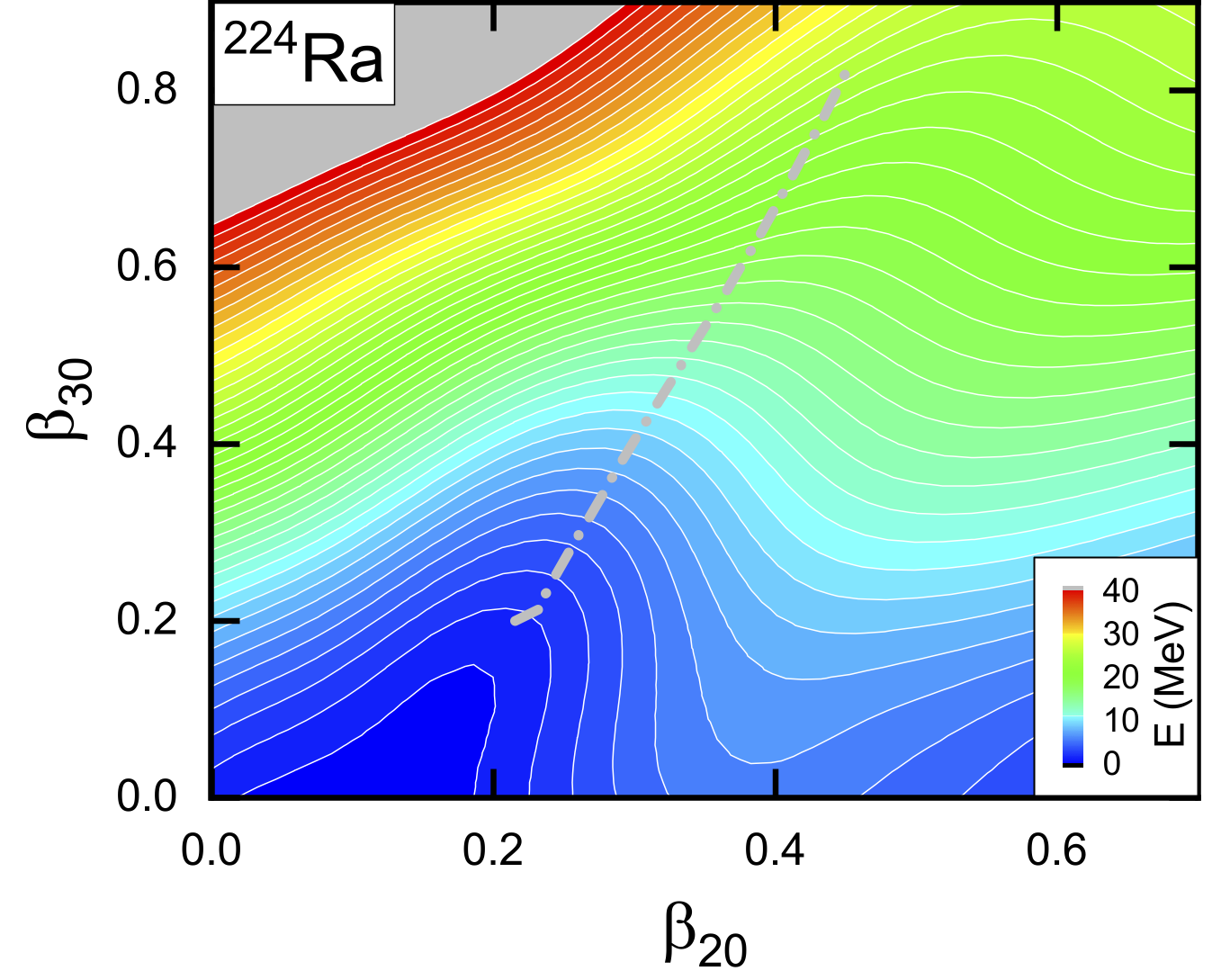}}
 \caption{Reflection symmetric deformation energy surface
of  $^{224}$Ra in the quadrupole-octupole axially-symmetric
plane. The grey curve denotes the dynamical
(least-action) path for $^{14}$C emission from equilibrium
deformation to scission.}
 \label{fig:224Raclus}
\end{figure}

As an example of cluster decay, we consider the emission of $^{14}$C from $^{222}$Ra and $^{224}$Ra.  
Cluster radioactivity was previously explored in a study based on least-action calculations with the Gogny energy density functional \cite{war11}. Fig. \ref{fig:224Raclus} displays the reflection asymmetric deformation energy surface
of  $^{224}$Ra in the quadrupole-octupole axially-symmetric plane. The grey curve denotes the dynamical
least-action path for $^{14}$C emission from equilibrium deformation to scission.
The scission point is located at ($\beta_{20}=0.45, \beta_{30}=0.82$). The corresponding contribution to the dimensionless action integral (\ref{eq:act_integration}) is 13.97, and from the scission to the outer turning point the  action is 27.95. The corresponding cluster half-live log$_{10}$ $\left(T_{^{14}C}\left[s\right]\right)=15.87$, is in excellent agreement with the experimental value log$_{10}$ $\left(T_{exp}\left[s\right]\right)=15.86$. In a similar calculation with the Gogny functional, the value log$_{10}$ $\left(T_{^{14}C}\left[s\right]\right)=15.06$ (cf. Fig. 13 of \cite{war11}) was obtained 
for the $^{14}$C decay half-life of $^{224}$Ra.

\begin{figure}[tbh!]
\scalebox{0.30}{\includegraphics{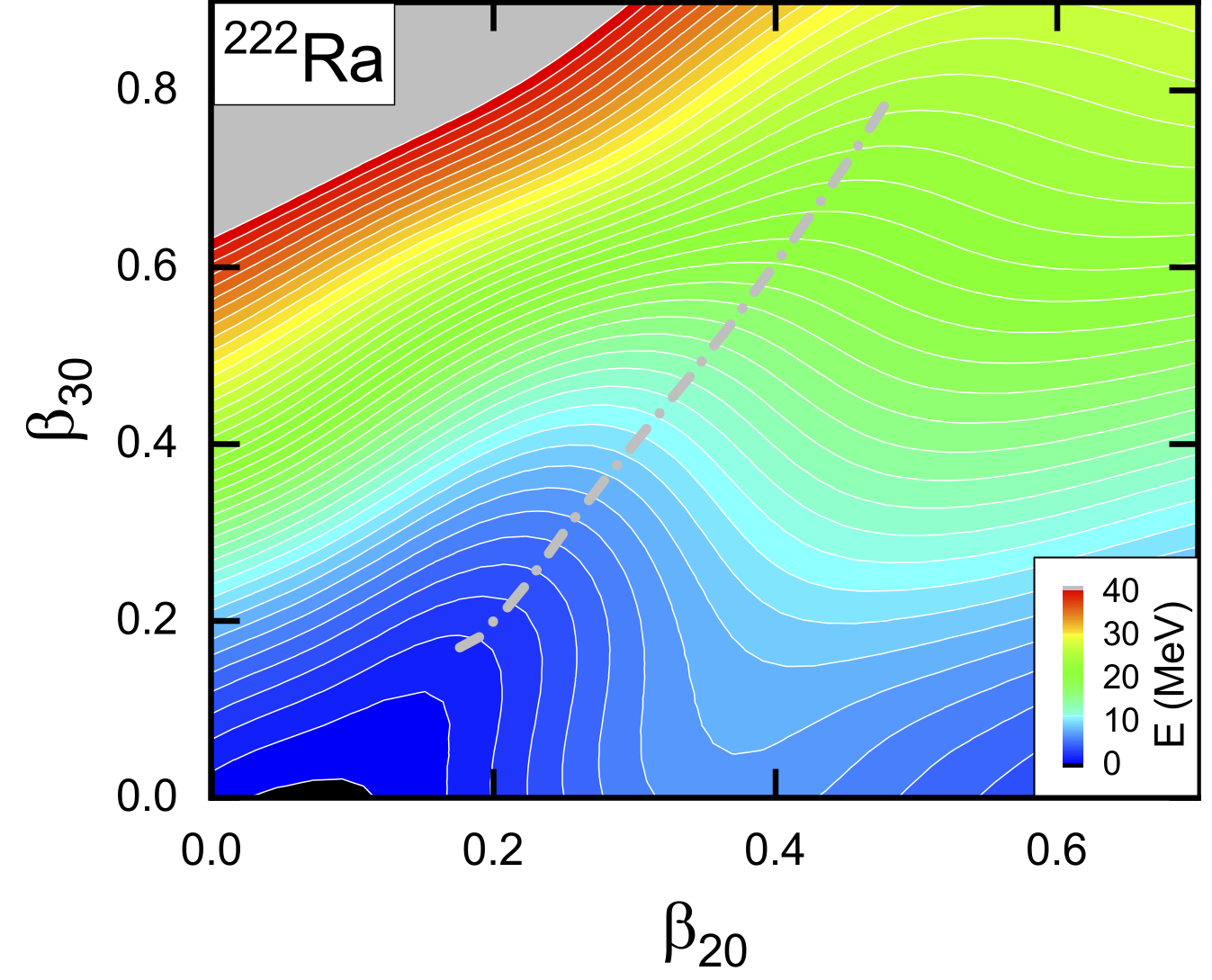}}
 \caption{ Same as in the caption to Fig. \ref{fig:224Raclus} but for $^{222}$Ra.}
 \label{fig:222Raclus}
\end{figure}

In Fig. \ref{fig:222Raclus} we show the corresponding deformation energy surface of $^{222}$Ra, and the least action path followed till the scission point, located at ($\beta_{20}=0.48, \beta_{30}=0.79$),  close to the value obtained for $^{224}$Ra. The corresponding contribution to the dimensionless action (\ref{eq:act_integration}) is 15.28 up to the scission point, and 24.04 from the scission point to the outer turning point. The predicted cluster half-live is log$_{10}$ $\left(T_{^{14}C}\left[s\right]\right)=13.61$, to be compared with the experimental value log$_{10}$ $\left(T_{exp}\left[s\right]\right)=11.01$. This appears to be on the same level of agreement with experiment as the result obtained using the Gogny functional: log$_{10}$ $\left(T_{^{14}C}\left[s\right]\right)=8.9$ \cite{war11}. The energy surfaces and the least-action paths for $^{14}$C emission are very similar for $^{222}$Ra and $^{224}$Ra. A Q-value that is 3 MeV larger in the case of $^{222}$Ra, explains its smaller action from the scission point to emission, and correspondingly its shorter cluster emission half-life.  

\begin{figure}[tbh!]
\scalebox{0.206}{\includegraphics{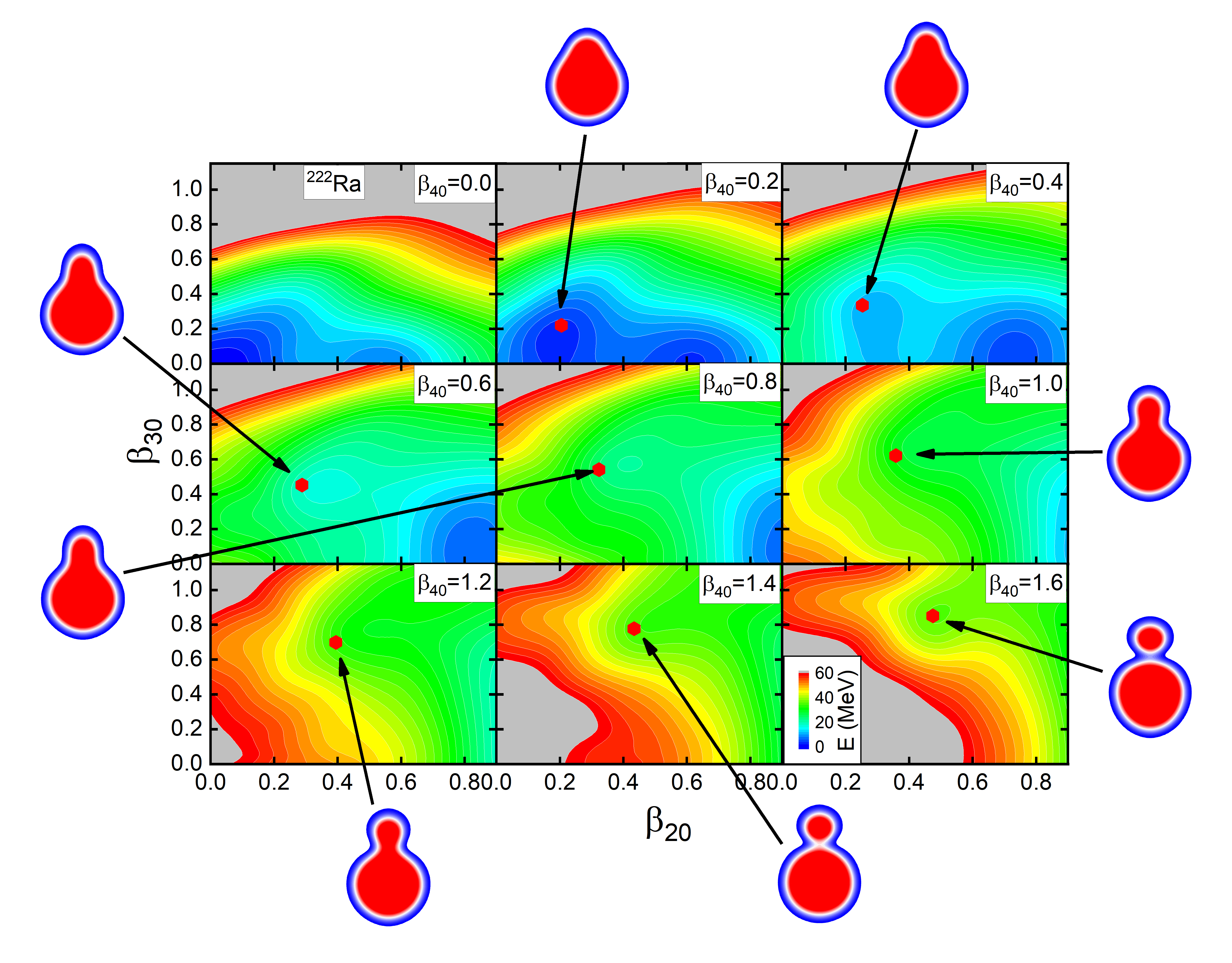}}
 \caption{Deformation energy surface of $^{222}$Ra in the
quadrupole-octupole axially-symmetric plane, for selected
values of the hexadecapole deformation $\beta_{40}$. Contours join
points on the surface with the same energy and red circles
indicate the points on the dynamical (least-action) path for $^{14}$C 
 emission.
 The insets display the intrinsic nucleon densities at selected points on the dynamical path.}
 \label{fig:222Raclus3D}
\end{figure}

Just as in the case of $\alpha$-decay, the hexadecupole degree of freedom could also play an important role for heavier cluster emission, such as $^{14}$C. We have thus carried out 3D hexadecapole-octupole-quadrupole calculations  of $^{14}$C emission from $^{222}$Ra. Fig. \ref{fig:222Raclus3D} displays the projections of the deformation energy surface onto the quadrupole-octupole plane, for several values of the hexadecapole parameter $\beta_{40}$. Compared to $\alpha$ emission, the deformation parameters reach much larger values: the scission point on the surface ($\beta_{20}$, $\beta_{30}$, $\beta_{40}$) is at ($0.49, 0.87, 1.66$). The least-action integral up to the scission point is 18.24, and the one form the scission point to the outer turning point is 22.48. The predicted half-life is log$_{10}$ $\left(T_{^{14}C}\left[s\right]\right)=14.82$. Hence, the inclusion of the hexadecapole degree of freedom produces a limited effect, with an increase of about one order of magnitude for the half-life. This increase is mainly due to the different locations of the scission point in 2D and 3D calculations. 

\begin{figure}[tbh!]
\scalebox{0.27}{\includegraphics{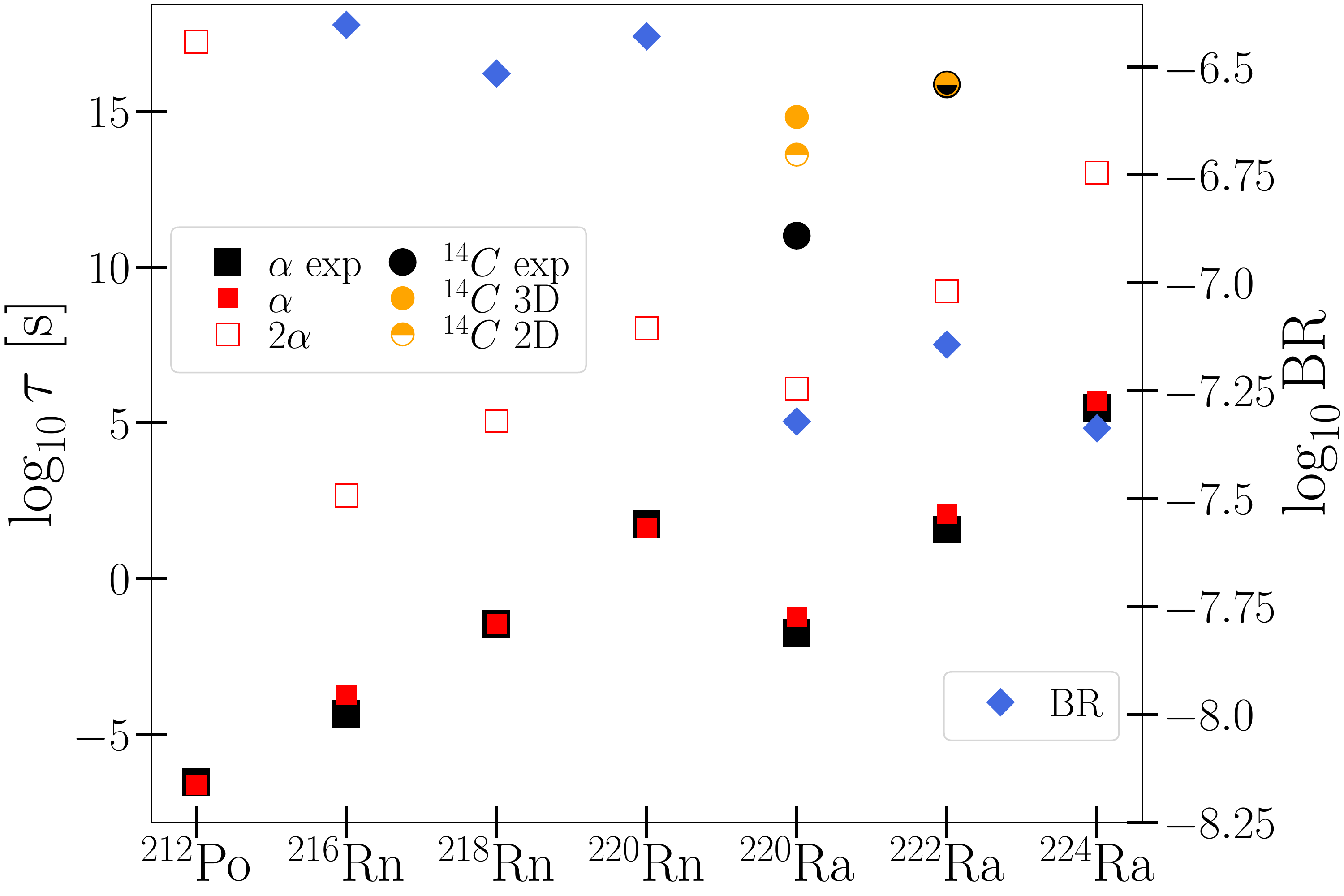}}
 \caption{Theoretical results for the $\alpha$, $2 \alpha$ and cluster decay half-lives of $^{216-220}$Rn, $^{220-224}$Ra and $^{212}$Po, in comparison with available experimental values.}    
 \label{fig:tot}
\end{figure}

In the case of heavier cluster emission, a better agreement with data could possibly be obtained by improving the treatment of collective inertia. It would be interesting to explore the effect of the perturbative and nonperturbative cranking approximations for the multidimensional inertia tensor, on the calculated half-lives for cluster decay. However, such an analysis is numerically rather involved and beyond the scope of the present study. Table \ref{tab:clust} summarizes the predicted $^{14}$C half-lives, and Fig \ref{fig:tot} presents an overview of the theoretical results for the half-lives of  $^{216-220}$Rn, $^{220-224}$Ra and $^{212}$Po, in comparison with available data. The overall good agreement with the data, especially in the $\alpha$-decay case, shows that the present approach provides a consistent microscopic framework for calculating $\alpha$ and cluster emission lifetimes, taking explicitly into account the dynamical path on a multidimensional deformation energy surface, the effective collective mass, and the pairing interaction. 

\setlength{\tabcolsep}{12pt}
\begin{center}
\begin{table}[h]
\caption{\label{tab:clust}}
Experimental, 2D and 3D calculated log$_{10}$ T[s] of the half-lives for $^{14}$C cluster emission in the case $^{222}$Ra and $^{224}$Ra.\\
\begin{tabular}{cccccc}
  & T$_{exp}$&   T$^{2D}_{C} $ & T$^{3D}_{C}$
 \\ \hline
 $^{222}$Ra & 11.01 &    13.61  & 14.82
\\ 
 $^{224}$Ra & 15.86 &   15.87   &  
\\
 \end{tabular}
\end{table}
\end{center}
\setlength{\tabcolsep}{6pt}

\section{\label{sec6} Summary}

$\alpha$, 2-$\alpha$ and $^{14}$C spontaneous emissions from heavy nuclei have been explored using a microscopic framework, based on the RHB model with the DD-PC1 functional and a separable pairing force. The corresponding 3D deformation energy surfaces (quadrupole, octupole, and hexadecapole collective degrees of freedom) enable the calculation of least-action integrals along dynamical paths. The predicted half-lives for $\alpha$-decay are within one order of magnitude of the experimental values. The predictions for heavier cluster emission deviate from the experimental values by 1 to 4 orders of magnitude. This may point to a possible improvement in the calculation of the effective collective inertia in the latter case. 

Two main effects impact the decay half-life: the dynamical path from the inner turning point up to the scission point and, 
for the part of the path beyond the scission point, the Q-value of the transition.
The relative importance of these two effects has been analyzed. 
$2 \alpha$ symmetric decay has been predicted within the same framework. All nuclei considered in the present study are also predicted to be $2 \alpha$ emitters. The corresponding branching ratio is more favorable for the Rn isotopes, but the Ra nuclei appear also as good candidates for the experimental observation of this rare decay mode. This has been confirmed in a recent phenomenological study of the double $\alpha$-decay half-life \cite{den22}.

\begin{acknowledgments} 
This work has been supported in part by the QuantiXLie Centre of Excellence, a project co-financed by the Croatian Government and European Union through the European Regional Development Fund - the Competitiveness and Cohesion Operational Programme (KK.01.1.1.01), and by the Institut Universitaire de France. J. Z. acknowledges support by the National Natural Science Foundation of China under Grant No. 12005107 and No. 11790325. 
\end{acknowledgments}

\end{document}